\documentclass[preprint,journal=jctcce]{achemso}
\usepackage{amssymb}
\usepackage[fleqn]{amsmath}
\usepackage{graphicx} 
\usepackage{mathtools}
\usepackage{xcolor}
\usepackage[title]{appendix}
\usepackage[hidelinks,colorlinks=false]{hyperref}
\usepackage[capitalise]{cleveref}
\usepackage{url}
\usepackage{tikz}
\usetikzlibrary{backgrounds}
\usepackage[super]{natbib}
\setcitestyle{citesep={,}}
\usepackage[frozencache,cachedir=.]{minted} 
\usepackage{natmove}
\usepackage{longtable}

\renewcommand{\emptyset}{\ensuremath{\{\}}}

 \def\sub{(0,0) circle (2cm)}
\def\mid{(-60:2cm) circle (2cm)}
\def\pred{(0:2cm) circle (2cm)}

\newcommand{\Id}{\ensuremath{I\! d}}

\title{Tensor Algebra Processing Primitives (TAPP):\\ Towards a Standard for Tensor Operations}
\author{Jan Brandejs}
\affiliation{Laboratoire de Chimie et Physique Quantique, UMR 5626 CNRS — Université de Toulouse, 118 route de Narbonne, F-31062 Toulouse, France}
\email{
jbrandejs@irsamc.ups-tlse.fr
}
\author{Niklas H\"ornblad}
\affiliation{Department of Computing Science, Umeå University, Umeå, 901 87, Sweden}
\author{Edward F. Valeev}
\affiliation{Department of Chemistry, Virginia Tech, Blacksburg, VA 24061, USA}
\author{Alexander Heinecke}
\affiliation{Intel Advanced Technologies Group,
Intel Corporation, Santa Clara, USA.}
\author{Jeff Hammond}
\affiliation{NVIDIA, 2788 San Tomas Expressway, Santa Clara, CA 95051}
\author{Devin Matthews}
\affiliation{Department of Chemistry, Southern Methodist University, Dallas, TX 75275, USA}
\author{Paolo Bientinesi}
\affiliation{Department of Computing Science, Umeå University, Umeå, 901 87, Sweden}
\date{December 2025}

\begin{document}

\maketitle


\section{Introduction}



Tensor operations drive fields such as artificial intelligence (AI) and quantum science modeling. In these fields, the computationally most intensive operation is typically tensor contraction.
The performance of tensor contraction has a direct effect on advances in material science,\cite{Irmler2019} quantum chemistry,\cite{Calvin2015} drug discovery,\cite{Kozinsky2023} life sciences and more.\cite{Kolda2009} This is because in these application fields, the cost of tensor contraction determines the largest affordable model size. For certain methods in these fields, the improvement in performance of tensor contractions has been the cornerstone of success, for instance, in deep learning, quantum computing simulation techniques, tensor network methods, and many quantum chemistry methods. 
Nowadays, the importance and impact of tensor operations on the world at large is staggering, especially in relation to the explosive growth of AI technologies. As an example, it is expected that in the coming years the annual electricity spent on tensor contractions globally will far exceed the total consumption of Germany~\cite{IEA2025}.

Despite its critical role in such widespread applications, the world of tensor contraction software is not as mature and well organized as that of matrix operations.  
The diversity and rapidly increasing number of developers and users of tensor software leads to a noticeable fragmentation of algorithms and codes.
Indeed, a recent survey highlighted the existence of a large number of scattered libraries and a high degree of code duplication~\cite{Bientinesi2021}.
One of the main reasons for this is a lack of standardization of tensor operations primitives. Widely adopted standards for tensor operations would allow developers to modularize their code and to reuse libraries without worrying about hard-to-replace dependencies. Regrettably, such standards are missing.

Recently, a community of tensor software developers, application experts, and industry representatives gathered to address this issue.\cite{workshop2024,workshop2025}
Among participants, the consensus is that the state of the art in multilinear algebraic operations (tensor algebra) is mature enough to 
support the emergence of a standard set of primitive 
building blocks. This paper makes the first step in this direction: We carefully define the problem of contracting tensors, present a proposal for a standard interface, and  provide a reference implementation. 

\subsection{Evolution}

For over four decades, \textbf{BLAS} (Basic Linear Algebra Subroutines) has been the de facto standard interface for fundamental linear algebra operations. 
The initial specification, now referred to as Level 1 BLAS, was published in 1979 to ``support'' scalar and vector operations and was directly adopted by the LINPACK library~\cite{Lawson1979, Dongarra1979}.
Recognizing the performance limitations of vector-only operations on modern architectures with memory hierarchies, a specification for Level 2 BLAS (matrix-vector operations) was proposed in 1988~\cite{Dongarra1988}, quickly followed by the proposal for Level 3 BLAS (matrix-matrix operations) in 1990 and adopted by the LAPACK library\cite{Demmel1991} among others.
Compared to Level 1 and 2, level 3 BLAS offers a higher ratio of arithmetic operations to memory accesses (arithmetic intensity), allowing to hide the slowness of memory access—a critical feature as processors became faster than memory.~\cite{Dongarra1990}.

Two factors contributed to the initial success of the BLAS interface. On the one hand, fixing the interface made it possible for developers to deliver highly optimized implementations, enabling performance portability. 
On the other hand, BLAS came into spotlight thanks to high-profile application in LINPACK benchmark\cite{Dongarra1988a} and in LAPACK which relies heavily on high-intensity Level 3 operations. 
BLAS eventually became the universal language of matrix operations, and its status stayed largely unchallenged, despite missing support for tensor contractions.

The standardization 
of BLAS occurred organically; it started with a reference implementation that solved immediate problems (Level 1 for LINPACK) and the interface then iteratively expanded towards matrix operations to maximize arithmetic intensity.
The organic rise was driven by bottom-up community consensus and the practical necessity of key applications like LINPACK and LAPACK, which solidified the standard through widespread utility rather than a rigid, top-down mandate.

In other standardization efforts, the order of events was different. 
There were about 9 unsuccessful efforts to standardize distributed communication before MPI (Message Passing Interface) was established~\cite{Gropp1996}. Experts with different backgrounds from industry, academic and research institutions, came together as part of the MPI forum, and it was a multi-year effort for them to reach consensus on how to encapsulate elementary operations. 

In the case of standardization of intranode parallelization, the first massive effort, POSIX Threads (pthreads)\cite{Nichols1996} established a low-level standard interface for explicit thread management. 
However, pthreads provides a complex, low-level interface. For parallelizing HPC applications, the OpenMP specification\cite{chandra2001} emerged and gained broad adoption, which 
can be attributed to its high-level programming model of incremental parallelization through compiler directives. This suggests that the HPC community favors productive and accessible standards, even when a more powerful but complex alternative exists.

Learning from past and ongoing standardization efforts, we assess that one should not try to define several axiomatic rules and to artificially build a standard on top of them. 
The standardization itself is an organic evolution based on experience and community interaction and consensus, rather than a deterministic process which could be planned by an individual. A key observation is that the process is catalysed by dedicated discussions, sharing of experience and software-reuse within the community.

The influence of those software standards (BLAS, pthreads, OpenMP, MPI) on computing has been profound. Once BLAS became widely adopted as a way to use hardware more efficiently, in turn, this  affected the hardware design. Classical computational hardware was tuned to execute GEMM because Linpack was the benchmark used to assess, and more importantly, to compare the performance of machines~\cite{Dongarra2024}. The development of HPC hardware and software went hand-in-hand.
The software architecture in scientific computing has been driven by these standards too.
In quantum chemistry, for instance, the tensor contraction, a fundamental multi-linear operation, has long been implemented by mapping it onto matrix multiplication, a pattern often referred to as ``transpose-transpose-GEMM-transpose'' (TTGT)~\cite{Springer2018:554}.

While tensor contractions can be expressed in terms of matrix operations, this not advisable. 
While functional, the TTGT approach is often suboptimal due to
data reshaping overhead and inefficiency for certain contractions.
TTGT places the burden on the developer, who must manually orchestrate data layouts, and it can obscure opportunities for performance optimizations that a more direct implementation could exploit. Furthermore, the underlying multilinear structure of the operation is lost in the translation to matrix-matrix products.
This led to the development of dedicated tensor software which encapsulates the tensor operations in a way appropriate for relevant performance optimizations. In the following, we survey some of those efforts that represent important conceptual milestones.

\subsubsection*{Early efforts} 
Published in 1996, POOMA~\cite{Reynders1996} was the earliest C++ library to use expression templates on top of distributed multidimensional arrays, without tensor contraction support, though. FTensor~\cite{Landry2003}, first published in 1998, was
one of the earliest C++ tensor contraction libraries, also built on expression templates.
In 2003, the Tensor Contraction Engine (TCE)~\cite{Hirata2003} automated the process of generating efficient code for tensor contractions and notably pioneered the use of the Loop-over-GEMMs technique for distributed tensors, later explored for node-level-optimization~\cite{DiNapoli2014}.

\subsubsection*{Distributed memory} 
Cyclops Tensor Framework (CTF)~\cite{Solomonik2014} introduced a distributed memory approach, allowing for the contraction of very large tensors on parallel computing systems.
Elemental~\cite{Poulson2013}, developed concurrently with CTF represents a move towards creating a unified framework for high-performance linear algebra on distributed-memory systems, including some tensor operations but not the contraction. Elemental has since been extended to tensor contraction in the context of artificial intelligence (AI) in the DiHydrogen library.\cite{dihydrogen}
TiledArray~\cite{Calvin2015} supports block-sparse tensor computation in distributed-memory setting.

\subsubsection*{Hardware-centric design} 
Fastor~\cite{Poya2017} brought the attention to hardware-centric design, built from the ground up to perform contractions for tensors with size known at compile-time through Single Instruction Multiple Data (SIMD) vectorization.
Loops-over-GEMMs\cite{DiNapoli2014,Li2015} slices tensors into a sequence of 2D sub-tensors (matrices), and contract them
via GEMMs.
TBLIS\cite{Matthews2018} introduced a novel approach to tensor contractions that hides the cost of the transposition step just before the matrix multiply kernel is called by permuting the data during processor cache offload.
Simultaneously, the GETT transpose-free scheme~\cite{Springer2018:554} was demonstrated and notably later implemented for GPUs in
cuTENSOR~\cite{cutensor}, which provides a set of optimized primitives to other more complete tensor libraries. cuTENSOR uses an elaborate cost model to generate code optimal for a given contraction by just-in-time compilation and to fuse the transposition and contraction steps again by remapping subtensors before kernel calls.

\subsubsection*{High-level frameworks} 
While not a specialized tensor library, NumPy~\cite{Harris2020} introduced the \textit{ndarray} object, and its powerful indexing and broadcasting capabilities have made it the de facto standard for numerical computing in Python. The \textit{einsum} function operating on this object is becoming  widely adopted and presents a clear candidate as the tensor contraction interface within the Python community.
TensorFlow~\cite{tensorflow} and PyTorch~\cite{pytorch}, developed for machine learning, have stood out by bringing high-performance tensor computations to a much wider audience than before.


\subsubsection*{Tensor compilers} 
Foundational ideas of tensor compilers materialized in Halide~\cite{RaganKelley2013}, particularly decoupling the algorithm from the schedule. Tensor Algebra Compiler (TACO)~\cite{Kjolstad2017} was the first tensor compiler to handle any tensor algebra expression involving sparse tensors. Tensor Virtual Machine (TVM)~\cite{chen2018} addressed the key machine learning challenge of deployment---fragmentation and incompatibility---by creating a unified compilation stack. Additionally, all AI targeted DSLs (domain specific languages) are based on MLIR~\cite{9370308}.

\subsubsection*{Current state and outlook}
Current trends in tensor contraction library development include new features for sparse calculations, GPU acceleration, Domain-Specific Languages (DSL), tensor decompositions\cite{Kolda2009}, Tensor Networks (TN)~\cite{Fishman2022}, and tensor compilers dedicated to neural networks~\cite{Li2021,Lianmin2024}.

Besides the tensor software listed here, there have been numerous other efforts~\cite{Bientinesi2021}.  We would like to bring to the attention of the reader that a large number of tensor libraries have emerged, the vast majority of which were developed independently, redundantly, and sub-optimally~\cite{Bientinesi2021}. In light of such an overcrowded and disorganized software landscape, we call for a process of identification and standardization of tensor primitives as reusable modules on which the community can build. As a first step, we focus on tensor contractions.

\subsection{The Aim of TAPP}

We observe evolution and growth of the field of tensor software, but no convergence towards a common standard interface for tensor operations. With more researchers entering the field, we expect the fragmentation will grow even further. The only widely adopted interface is the Pythonic \texttt{einsum}, which, while designed for high-level scripting, does not generalize easily to compiled languages or for low-level code integration.

To address the missing standard, we define the Tensor Algebra Processing Primitives (TAPP), a
standard C-based interface for tensor operations.
The first aim of TAPP is to define an interface for tensor contraction operations, to decouple the application layer from the implementation layer.

We aim to specify a BLAS-like interface to correctly encapsulate tensor operations while allowing performance portability. 
The interface specification needs to be prepared hand in hand with a clear reference implementation.
By promoting specific tensor kernels to standardized building blocks, TAPP enables a clear separation of concerns:
\begin{enumerate}
\item Application developers can write code against the TAPP interface once, gaining access to multiple backends without having to change their code base once a need to switch the backend arises.
\item Library developers can optimize specific kernels (TAPP implementations) without needing to support every application framework.
\item Hardware vendors can provide TAPP-compliant libraries to ensure their hardware is accessible to the wider scientific community.
\end{enumerate}
This architecture seeks to mirror the success of BLAS: it creates a transparent market where performance can be measured and compared, and where optimizations in the backend can automatically propagate to all applications using the standard.

\subsection{The TAPP Ecosystem}

The TAPP initiative is built on a consensus among academic and industrial stakeholders, provides a suite of resources that allow a relatively simple integration and enable long-term viability.
Central to this ecosystem is the Reference Implementation, presented in this article. This implementation prioritizes readability and serves for correctness check during integration of TAPP. It allows hardware vendors and library developers to validate their optimized kernels against a known correct result on a comprehensive test set which includes peculiar edge cases. The source code of reference implementation is openly accessible on GitHub under a permissive BSD-3 license.\cite{tapporg}

The TAPP community aims to provide a growing set of high-performance backends. TAPP interfaces have already been developed for libraries such as TBLIS and cuTENSOR, demonstrating that the standard can support diverse architectures without changing the application code.

TAPP follows an open governance model similar to the evolution of MPI and OpenMP. It is maintained by a working group representing application experts, library developers, and hardware vendors who assembled at the CECAM Workshop on Tensor Contraction Library Standardization in Toulouse in May 2024 and at the 2nd Toulouse Tensor Workshop in September 2025. This ensures that the standard evolves based on practical feedback rather than imposing a rigid top-down mandate.

It follows from the above that TAPP itself is a broader effort with a series of outputs, meetings and communications beyond this article.
However, TAPP has just reached a key milestone with the finalization of the reference implementation and the purpose of this article is to present it to the wider community at that very moment. The reference implementation itself is a key goal of the entire TAPP initiative and the main focus of this article, because it clarifies the interface to the small details and elucidates most ambiguities and edge-cases, which presents a major challenge when trying to define the interface rigorously.

\subsection{Target Use Case}

A compelling example is found in the development of the DIRAC,\cite{dirac} a legacy Fortran quantum chemistry package. Originally a legacy code using direct BLAS calls, its performance-critical modules were modernized to rely on the ExaTENSOR library.\cite{Pototschnig2021} However, when the main developer of ExaTENSOR moved to industry, the library ceased to be maintained, leading to a roadblock in key applications over time. To mitigate this, the team was forced to rewrite the entire code base. This "dependency hell" exposed the urgent lack for a standard and led the DIRAC team to issue a call-to-arms to the community via the 2024 workshop on standardization.

If DIRAC had been written against the TAPP interface, switching from an unmaintained backend to a supported one (e.g., TBLIS or a vendor-provided library) would have been a link-time decision rather than a rewrite.

This scenario is not unique to quantum chemistry. For High-level Languages, TAPP provides a stable API, allowing frameworks like Julia,\cite{julia} PyTorch\cite{pytorch} or MATLAB\cite{MATLAB} to safely offload heavy contractions to whichever optimized library is available to the user.



\section{Mathematical Definitions} \label{sec:definitions}

Since the contraction of two tensors is the generalization of the matrix product, it is natural to extend the notation used for matrices to tensors. 
Specifically, following the BLAS interface, a matrix product is represented by the following \emph{ternary} operation:
\begin{equation} 
  \label{product}
  C \coloneqq \alpha A  B + \beta C,
\end{equation}
where $A, B$ and $C$ are matrices (of compatible size), and $\alpha$ and $\beta$ are scalars. The index-based notation unambiguously captures how the elements of $C$ are computed: 
\begin{equation}
    \forall i,j: \  C_{i,j} \coloneqq \beta C_{i,j} + \alpha \sum_k A_{i,k} B_{k,j}.
\end{equation}
The extension to tensors appears to be straightforward: given the tensors $A, B$ and $C$, the contraction of $A$ with $B$ and the update of $C$ is also represented by Eq.~\eqref{product}.
However, in contrast to the 2-dimensional case (matrices), this notation does not unequivocally capture how the elements of $C$ are computed for general tensor contractions. Indeed, the precise computation behind expression~\eqref{product} depends on the indices of the different tensors. 

In this section, we consider a generalized form of \cref{product}:
\begin{equation} 
  \label{binary_contraction}
  D \coloneqq \alpha A  B + \beta C,
\end{equation}
and provide concrete computational definitions for tensor contraction and variant operations that are supported by TAPP.

\subsection{Preliminaries}

Terminology for describing tensors and tensor operations varies widely across (and even within) various disciplines. We choose a ``standard'' set of terms here---see \cref{sec:glossary} for a glossary of synonyms and related terms. A tensor $T$ is characterized by the number of 
{\em indices}, by the {\em extent} of the  indices, and by the entries (data). 
Specifically, let $n$ be an integer, indicating the number of indices of $T$.
If $n > 0$, the extent of the $i$-th index,  with $i \in [1,\dots,n]$, is indicated by the positive integer $d_i$. 
We refer to the object
$T \in \mathbb{R}^{d_{1} \times d_{2} \times \dots \times d_{n}}$
as the {tensor} $T$, with \emph{shape} $\bigcup\limits_{i=1}^{n}\{d_i\}$.

\paragraph{Note 1:} Everything presented in this document applies to tensors that live in $\mathbb{C}$.
\paragraph{Note 2:} In this notation, scalars correspond to tensors with $n = 0$ indices, but can also be represented as tensors with all extents equal to 1.
\paragraph{Note 3:} This notation does not carry information about how the entries of $T$ are stored in memory. This means that any permutation of the indices identifies the same tensor. As an example, the same tensor $T$ can be defined both as 
$\mathbb{R}^{d_{1} \times d_{2} \times \dots \times d_{n}}$,
and as 
$\mathbb{R}^{d_{2} \times d_{1} \times \dots \times d_{n}}$ (indices are permuted). \\

For convenience, each index is normally assigned a label. 
\Id\ is the function that maps indices to labels. 
This function   satisfies the condition that 
$\Id(i) = \Id(j)$ only if $d_i = d_j$; this 
means that if two or more indices have equal extents, then they may, but do not have to, share the same label. 
The set of labels for tensor $T$ is denoted by $\mathcal{T}$. 
It is $\mathcal{T} \coloneqq \bigcup\limits_{i=1}^{n}\{\Id(i)\}$, and  $1 \le |\mathcal{T}| \leq n$.

\subsection{Operations}
\label{sec:preliminaries:operations}
Let $n > 0, p > 0, r > 0$ be integers, 
and consider the tensors
\[
A \in \mathbb{R}^{d_{1} \times d_{2} \times \dots \times d_{n}},\
B \in \mathbb{R}^{e_{1} \times e_{2} \times \dots \times e_{p}}, \text{\ and \ }
C, D \in \mathbb{R}^{f_{1} \times f_{2} \times \dots \times f_{r}}.  
\]

\paragraph{Note 4:} $n$, $p$ and/or $r$ could also be 0. In those cases, the operation~\eqref{binary_contraction} turns into the scaling of a tensor, the product of two scalars, or a tensor dot product. We focus on the non-degenerate case for clarity.

\paragraph{Note 5:} The indices of $C$ and $D$ only differ by permutation. In practice, implementation details restrict $C$ and $D$ to identical memory layouts as well. \\

Let 
$\mathcal{A, B, C, D}$ denote the set of labels for $A, B, C, D$, respectively, and let the integers $o >0,q >0,s > 0$ indicate the cardinality of $\mathcal{A, B, D}$, respectively: 
$o \coloneqq |\mathcal{A}|$, $q \coloneqq |\mathcal{B}|$, and $s \coloneqq |\mathcal{C}| = |\mathcal{D}|$.
The meaning of Eq.~\eqref{binary_contraction} varies according to the relationship between, and the cardinality of, the sets 
$\mathcal{A, B}$ and $\mathcal{D}$.
This was fixed by the decision that isolated indices in $\mathcal{D}$ will not be supported (see Sec.~\ref{sec:preliminaries:operations:case5}).
Consequently, in the following text we assume that every index in 
$\mathcal{D}$ appears also in $\mathcal{A}$ or in $\mathcal{B}$ (or in both), that is,
\begin{equation}
  \label{eq:FreeD}
  \mathcal{D} / (\mathcal{A} \cup \mathcal{B}) = \emptyset.
\end{equation}

\subsubsection{Case 1: ``Simple" Contractions
}
\label{sec:preliminaries:operations:case1}

This case is the higher dimensional counterpart of the matrix-matrix product. The indices in $\mathcal{A}$ are split into two disjoint groups, 
$\mathcal{P}$ and $\mathcal{F_A}$,
the \emph{contracted} and the \emph{free} indices, respectively. 
Similarly for $\mathcal{B}$: 
$\mathcal{P}$ and $\mathcal{F_B}$ contain
the contracted and free indices, respectively. 
Finally, $\mathcal{D}$  (and thus $\mathcal{C}$) only contains the free indices of $A$ and $B$: 
$\mathcal{D} \equiv \mathcal{F_A} \cup \mathcal{F_B}$.
Furthermore, no two indices in the same tensor share the same label (all index labels are distinct), i.e., 
$ n = o, \
  p = q, \
  r = s. $
\begin{figure}[!h]
  \begin{center}
    \begin{tabular}{l @{\hspace*{15mm}}r}
      \scalebox{.5}{
        \begin{tikzpicture}[thick]
          \begin{scope}[opacity=0.5]

            \draw[color=red, fill=red] \sub;
            \draw[color=blue, fill=blue] \pred;
            \draw[color=green, fill=green] \mid;
          \end{scope}
          \begin{scope}
            \draw[color=black] \sub;
            \draw[color=black] \pred;
            \draw[color=black] \mid;

            \draw (-2.5,-0) node {$A$};
            \draw (1,-4.2) node {$D$};
            \draw (4.5,0) node {$B$};
            
            \draw (-0.7,0.4) node {$\emptyset$};
            \draw (1,0.6) node {$\mathcal{P}$};
            \draw (2.7,0.4) node {$\emptyset$};
            \draw (0,-1.2) node {$\mathcal{F_A}$};
            \draw (1,-0.6) node {$\emptyset$};
            \draw (2,-1.2) node {$\mathcal{F_B}$};
            \draw (1,-2.4) node {$\emptyset$};
          \end{scope}
        \end{tikzpicture}
      }
      &
      \scalebox{.5}{
        \begin{tikzpicture}[thick]
          \begin{scope}[opacity=0.5]

            \draw[color=red, fill=red] \sub;
            \draw[color=blue, fill=blue] \pred;
            \draw[color=green, fill=green] \mid;
          \end{scope}
          \begin{scope}
            \draw[color=black] \sub;
            \draw[color=black] \pred;
            \draw[color=black] \mid;

            \draw (-2.5,-0) node {$A$};
            \draw (1,-4.2) node {$D$};
            \draw (4.5,0) node {$B$};
            
            \draw (-0.7,0.4) node {$\emptyset$};
            \draw (1,0.6) node {$\mathcal{P}$};
            \draw (2.7,0.4) node {$\emptyset$};
            \draw (0,-1.2) node {$\mathcal{F_A}$};
            \draw (1,-0.6) node {$\mathcal{H}$};
            \draw (2,-1.2) node {$\mathcal{F_B}$};
            \draw (1,-2.4) node {$\emptyset$};
          \end{scope}
        \end{tikzpicture}
      }
    \end{tabular}
  \end{center}
  \caption{Relations among the label sets $\mathcal{A}, \mathcal{B}$ and $\mathcal{D}$. Left: Simple contractions (Case 1); right: simple contractions and Hadamard products (Case 2).}
  \label{Fig:Simple-and-Hadamard}
\end{figure}
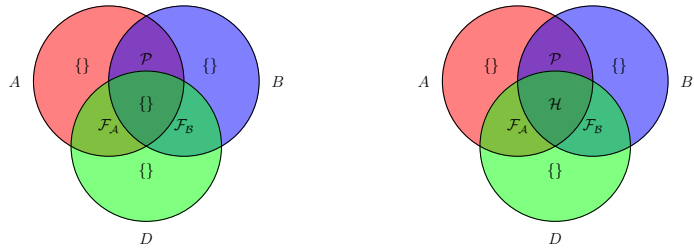

Under these conditions, 
the product~\eqref{binary_contraction} is computed by as many for loops as free indices ($\mathcal{D} \equiv \mathcal{F_A} \cup \mathcal{F_B}$), and a summation over the contracted indices ($\mathcal{P} \equiv \mathcal{A} \cap \mathcal{B}$):
\begin{equation} 
  D_{\mathcal{F_A} \cup \mathcal{F_B}} \coloneqq
  \alpha \cdot \sum\limits_{\mathcal{P}}A_{\mathcal{P} \cup \mathcal{F_A}} B_{\mathcal{P} \cup \mathcal{F_B}} +
  \beta \cdot C_{\mathcal{F_A} \cup \mathcal{F_B}}.
  \label{case1}
\end{equation}

\paragraph{Example} The contraction 
$D_{\delta \gamma} \coloneqq A_{\alpha \beta \gamma} B_{\beta \delta \alpha} + C_{\delta \gamma}$
is computed as
\begin{equation}
    \forall i \ \forall j \  D_{i,j} \coloneqq 
    \alpha \sum_k \sum_l A_{k,l,j} B_{l,i,k} + \beta C_{i,j}\ ,
\end{equation}
where 
$i \in [1,\dots,\text{extent(}\delta)]$,
$j \in [1,\dots,\text{extent(}\gamma)]$, 
$k \in [1,\dots,\text{extent(}\alpha)]$, and\break
$l \in [1,\dots,\text{extent(}\beta)]$.

\subsubsection{Case 2: Hadamard Product and Contraction}
\label{sec:preliminaries:operations:case2}

In this second case, we maintain the constraint that no two indices in the same tensor share the same label  (i.e., 
$ n = o, \
  p = q, \
  r = s.
  $), but we now allow one or more indices to appear simultaneously in $\mathcal{A}, \mathcal{B}$, and $\mathcal{D}$. This set of indices ($\mathcal{H} \coloneqq \mathcal{A} \cap \mathcal{B} \cap \mathcal{D}$) indicates an element-wise multiplication, also known as Hadamard product. 

  Concretely, the indices in $\mathcal{A}$, $\mathcal{B}$, and $\mathcal{D}$ are split into three disjoint sets: 
  $\mathcal{A} \equiv \mathcal{P} \cup \mathcal{F_A} \cup \mathcal{H}$, 
  $\mathcal{B} \equiv \mathcal{P} \cup \mathcal{F_B} \cup \mathcal{H}$, and
  $\mathcal{D} \equiv \mathcal{H} \cup \mathcal{F_A} \cup \mathcal{F_B}$, respectively.
  The product~\eqref{binary_contraction} is computed element-wise for all indices in $\mathcal{H}$, and
  otherwise as a ``case-1 contraction'' over the indices in $\mathcal{P}$:
\begin{equation}
  \forall \mathcal{H} :
  D_{\mathcal{H} \cup \mathcal{F_A} \cup \mathcal{F_B}} \coloneqq
  \alpha \cdot \sum\limits_{\mathcal{P}}
  A_{\mathcal{H} \cup \mathcal{P} \cup \mathcal{F_A}}
  B_{\mathcal{H} \cup \mathcal{P} \cup \mathcal{F_B}} +
  \beta \cdot
  C_{\mathcal{H} \cup \mathcal{F_A} \cup \mathcal{F_B}}.
 \label{case2}
\end{equation}

\paragraph{Example} The contraction 
$D_{\delta \alpha} \coloneqq A_{\alpha \beta} B_{\beta \delta \alpha} + C_{\delta \alpha}$
is computed as
\begin{equation}
    \forall i \ \forall j \  D_{j,i} \coloneqq 
    \alpha \sum_k A_{i,k} B_{k,j,i} + \beta C_{j,i}\ ,
\end{equation}
where 
$i \in [1,\dots,\text{extent(}\delta)]$, 
$j \in [1,\dots,\text{extent(}\alpha)]$, and
$k \in [1,\dots,\text{extent(}\beta)]$.

\subsubsection{Case 3: Repeated Indices}
\label{sec:preliminaries:operations:case3}

As for Case 2, 
the indices in $\mathcal{A}$, $\mathcal{B}$, and $\mathcal{D}$ are split into three disjoint sets: 
$\mathcal{A} \equiv \mathcal{P} \cup \mathcal{F_A} \cup \mathcal{H}$, 
$\mathcal{B} \equiv \mathcal{P} \cup \mathcal{F_B} \cup \mathcal{H}$, and
$\mathcal{D} \equiv \mathcal{H} \cup \mathcal{F_A} \cup \mathcal{F_B}$, respectively.
However, in contrast to Case 2, now it is possible for indices with the same extent to share the same label, that is, 
$ n \le o, \
  p \le q, \ \text{and} \ 
  r \le s$, where 
  $n, p$, and $r$ are the number of indices of $A, B$, and $D$, respectively, and 
  $o \coloneqq |\mathcal{A}|$, $q \coloneqq |\mathcal{B}|$, and $s \coloneqq |\mathcal{D}|$ are the number of distinct labels in $\mathcal{A}$, $\mathcal{B}$, and $\mathcal{D}$, respectively. 
  Computationally, this case is equivalent to Case 2 (Eq.~\eqref{case2}), with as many loops as unique indices.  
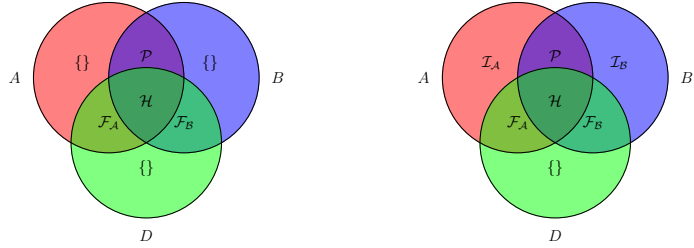
\begin{figure}[!h]
  \begin{center}
    \begin{tabular}{l @{\hspace*{15mm}}r}
      \scalebox{.5}{
        \begin{tikzpicture}[thick] 
          \begin{scope}[opacity=0.5]

            \draw[color=red, fill=red] \sub;
            \draw[color=blue, fill=blue] \pred;
            \draw[color=green, fill=green] \mid;
          \end{scope}
          \begin{scope}
            \draw[color=black] \sub;
            \draw[color=black] \pred;
            \draw[color=black] \mid;

            \draw (-2.5,-0) node {$A$};
            \draw (1,-4.2) node {$D$};
            \draw (4.5,0) node {$B$};
            
            \draw (-0.7,0.4) node {$\emptyset$};
            \draw (1,0.6) node {$\mathcal{P}$};
            \draw (2.7,0.4) node {$\emptyset$};
            \draw (0,-1.2) node {$\mathcal{F_A}$};
            \draw (1,-0.6) node {$\mathcal{H}$};
            \draw (2,-1.2) node {$\mathcal{F_B}$};
            \draw (1,-2.4) node {$\emptyset$};
          \end{scope}
        \end{tikzpicture}
      }
      &
      \scalebox{.5}{
        \begin{tikzpicture}[thick]
          \begin{scope}[opacity=0.5]

            \draw[color=red, fill=red] \sub;
            \draw[color=blue, fill=blue] \pred;
            \draw[color=green, fill=green] \mid;
          \end{scope}
          \begin{scope}
            \draw[color=black] \sub;
            \draw[color=black] \pred;
            \draw[color=black] \mid;

            \draw (-2.5,-0) node {$A$};
            \draw (1,-4.2) node {$D$};
            \draw (4.5,0) node {$B$};
            
            \draw (-0.7,0.4) node {$\mathcal{I_A}$};
            \draw (1,0.6) node {$\mathcal{P}$};
            \draw (2.7,0.4) node {$\mathcal{I_B}$};
            \draw (0,-1.2) node {$\mathcal{F_A}$};
            \draw (1,-0.6) node {$\mathcal{H}$};
            \draw (2,-1.2) node {$\mathcal{F_B}$};
            \draw (1,-2.4) node {$\emptyset$};
          \end{scope}
        \end{tikzpicture}
      }
    \end{tabular}
  \end{center}
  \caption{Relations among the label sets $\mathcal{A}, \mathcal{B}$ and $\mathcal{D}$. Left: Repeated indices (Case 3); right: Isolated indices (Case 4).}
  \label{Fig:repeat-and-isolated}
\end{figure}


\paragraph{Example} The contraction 
$D_{\delta \alpha} \coloneqq A_{\beta \alpha \beta} B_{\beta \delta \alpha} + C_{\delta \alpha}$
is computed as
\begin{equation}
    \forall i \ \forall j \  D_{j,i} \coloneqq 
    \alpha \sum_k A_{k,i,k} B_{k,j,i} + \beta C_{j,i}\ ,
\end{equation}
where 
$i \in [1,\dots,\text{extent(}\delta)]$, 
$j \in [1,\dots,\text{extent(}\alpha)]$, and
$k \in [1,\dots,\text{extent(}\beta)]$.

\paragraph{Note 6:} A tensor $A$ with a \emph{strided layout} and repeated index labels can equivalently be represented as a tensor $A'$ with unique labels (and hence fewer indices), with the relationship that the \emph{stride} of each index in $A'$ is equal to the sum of the strides of all indices in $A$ with the same label.

\subsubsection{Case 4: Isolated Indices in $\mathcal{A}$ or $\mathcal{B}$}
\label{sec:preliminaries:operations:case4}

This case is more general than the previous ones in that it is now possible that some labels only exist in  
$\mathcal{A}$, or only in $\mathcal{B}$: 
$\mathcal{I_A} \coloneqq \mathcal{A}/(\mathcal{B} \cup \mathcal{D})$, and 
$\mathcal{I_B} \coloneqq \mathcal{B}/(\mathcal{A} \cup \mathcal{D})$. 
This means that the indices in $\mathcal{A}$ and $\mathcal{B}$ are split into four disjoint sets,
$\mathcal{A} \equiv \mathcal{H} \cup \mathcal{F_A} \cup \mathcal{P} \cup \mathcal{I_A}$, and  
$\mathcal{B} \equiv \mathcal{H} \cup \mathcal{F_B} \cup \mathcal{P} \cup \mathcal{I_B}$, while the indices in $\mathcal{D}$ are split into three disjoint sets:
$\mathcal{D} \equiv \mathcal{H} \cup \mathcal{F_A} \cup \mathcal{F_B}$.

The indices in $\mathcal{I_A}$ ($\mathcal{I_B}$) indicate a \emph{reduction} within tensor $A$ ($B$). Computationally,
\begin{equation}
\forall \mathcal{H} : D_{\mathcal{H} \cup \mathcal{F_A} \cup \mathcal{F_B}} \coloneqq
\alpha \cdot 
\sum_{\mathcal{P}}
\left(\sum_{\mathcal{I_A}} A_{\mathcal{H} \cup \mathcal{F_A} \cup \mathcal{P} \cup \mathcal{I_A}}\right)
\left(\sum_{\mathcal{I_B}} B_{\mathcal{H} \cup \mathcal{F_B} \cup \mathcal{P} \cup \mathcal{I_B}}
\right)
+ \beta \cdot C_{\mathcal{H} \cup \mathcal{F_A} \cup \mathcal{F_B}}
\end{equation}.

\paragraph{Example} 
The contraction 
$D_{\delta \alpha} \coloneqq A_{\beta \alpha \beta} B_{\beta \gamma \delta \gamma \alpha} + C_{\delta \alpha}$
is computed as
\begin{equation}
    \forall i \ \forall j \  D_{j,i} \coloneqq 
    \alpha \sum_k A_{k,i,k} 
    \left( \sum_l B_{k,l,j,l,i} \right) + \beta C_{j,i}\ ,
\end{equation}
where 
$i \in [1,\dots,\text{extent(}\delta)]$, 
$j \in [1,\dots,\text{extent(}\alpha)]$, 
$k \in [1,\dots,\text{extent(}\beta)]$, and\break
$l \in [1,\dots,\text{extent(}\gamma)]$.

\subsubsection{Case 5: Isolated Indices in $\mathcal{D}$}\label{sec:preliminaries:operations:case5}

In the most general case, some labels may also only exist in $\mathcal{C}$ and hence $\mathcal{D}$: 
$\mathcal{I_C} \coloneqq \mathcal{C}/(\mathcal{A} \cup \mathcal{B})$.
This means that the indices in $\mathcal{C}$ and $\mathcal{D}$ are also split into four disjoint sets,
$\mathcal{C} = \mathcal{D} \equiv \mathcal{H} \cup \mathcal{F_A} \cup \mathcal{F_B} \cup \mathcal{I_C}$, bringing the total number of disjoint sets to seven: $\mathcal{P}$, $\mathcal{F_A}$, $\mathcal{F_B}$, $\mathcal{H}$, $\mathcal{I_A}$, $\mathcal{I_B}$, and $\mathcal{I_C}$, corresponding to all seven Venn diagram regions in Figures~\ref{Fig:Simple-and-Hadamard} and \ref{Fig:repeat-and-isolated}.

The indices in $\mathcal{I_C}$ indicate a \emph{broadcast} of elements from the product of $A$ and $B$ into multiple (presumably unique) locations in $D$.
Computationally, 
\begin{equation}
\forall \mathcal{I_C} \ \forall \mathcal{H} : D_{\mathcal{H} \cup \mathcal{F_A} \cup \mathcal{F_B} \cup \mathcal{I_C}} \coloneqq
\alpha \cdot 
\sum_{\mathcal{P}}
\left(\sum_{\mathcal{I_A}} A_{\mathcal{H} \cup \mathcal{F_A} \cup \mathcal{P} \cup \mathcal{I_A}}\right)
\left(\sum_{\mathcal{I_B}} B_{\mathcal{H} \cup \mathcal{F_B} \cup \mathcal{P} \cup \mathcal{I_B}}
\right)
+ \beta \cdot C_{\mathcal{H} \cup \mathcal{F_A} \cup \mathcal{F_B} \cup \mathcal{I_C}}.
\end{equation}

\paragraph{Example} 
The contraction 
$D_{\delta \alpha \epsilon} \coloneqq A_{\beta \alpha \beta} B_{\beta \gamma \delta \gamma \alpha} + C_{\delta \alpha \epsilon}$
is computed as
\begin{equation}
    \forall m \ \forall i \ \forall j \  D_{j,i,m} \coloneqq 
    \alpha \sum_k A_{k,i,k} 
    \left( \sum_l B_{k,l,j,l,i} \right) + \beta C_{j,i,m}\ ,
\end{equation}
where 
$i \in [1,\dots,\text{extent(}\delta)]$, 
$j \in [1,\dots,\text{extent(}\alpha)]$, 
$k \in [1,\dots,\text{extent(}\beta)]$,
$l \in [1,\dots,\text{extent(}\gamma)]$, and
$m \in [1,\dots,\text{extent(}\epsilon)]$.

\paragraph{Note 7:} The (partial) reduction of input tensors (Case 4) and the broadcasting into output tensors (Case 5) potentially imply either redundant computation or the need for unbounded amounts of workspace. In order to guarantee well-defined, near-optimal behavior, TAPP currently does not require the support for Cases~4 and 5.

\subsubsection{Unary and Binary Operations}
In analogy to contractions (trinary tensor operations), binary operations of the form $C \coloneqq \alpha A + \beta B$ can be viewed in terms of index label strings and disjoint subsets of labels based on their appearance in $A$ and/or $B/C$ (similarly to contraction, it must be the case that $\mathcal{B}=\mathcal{C}$): $\mathcal{A}\equiv\mathcal{H}\cup\mathcal{I_A}$ and $\mathcal{B}=\mathcal{C}\equiv\mathcal{H}\cup\mathcal{I_B}$. As before, indices in $\mathcal{I_A}$ imply reduction, while indices in $\mathcal{I_B}$ imply broadcasting. Finally, unary operations such as tensor scaling ($B\coloneqq\alpha A$) do not have distinct subsets of labels, but repeated labels may indicate operations on (semi-)\break diagonal elements of the tensor. While less important for high-performance computation, TAPP includes interfaces for binary and unary operations for convenience and consistency.

\section{The Interface}

\subsection{Related Work}

As for standardization of tensor operations, there has been some work in the past, but none of the interfaces have prevailed. 
BTAS\cite{btas,Shi2016} has aimed at providing a standard and a basic CPU implementation. 
During the Dagstuhl tensor workshop\cite{Dagrep2022} in 2022, a development of domain specific tensor language has been initiated, but remains unfinished and undocumented. 
There has been work technical specifications of additional aspects of standardization, like tensor memory distribution\cite{Valeev2023, cppref} and randomized multilinear algebra.\cite{Murray2023}
Last but not least, there have been discussions on standardization between the authors of TBLIS and cuTENSOR. One of their draft interfaces eventually served as a starting draft during the practical discussions leading to TAPP.

\subsection{Design Decisions: Possible Approaches} \label{sec:design-goals}

There are two common, high-level ways to encode tensor algebra:
\begin{itemize}
\item {\bf Tensor eDSL} (embedded Domain Specific Language):  More general math-like interfaces are only possible as an API in languages that support operator overloading, which allows to \emph{embed} a domain-specific language (DSL) for tensor algebra.
\item {\bf Einsum}: The \texttt{einsum} method is a popular way to encode evaluation of a single \emph{tensor network}, which is a more general contraction of many tensors (beyond the trinary operations discussed above). The advantage of such an interface is that it can be implemented in any language that supports functions.
\end{itemize}

ITensor, CTF, and TiledArray are examples of tensor frameworks that provide an eDSL. The main difference among these frameworks is that ITensor binds indices to tensors permanently (at the construction time).
For example, the following instance of \cref{binary_contraction},
\begin{align}
  D_{il} \coloneqq \alpha A_{ijk} B_{jlk} + \beta C_{li}
  \label{eq:example_DABC}
\end{align}
can be encoded in ITensor as,
\begin{minted}[frame=lines, linenos]{cpp}
    auto i = itensor::Index("i", i_len), j = itensor::Index("j", j_len);
    auto k = itensor::Index("k", k_len), l = itensor::Index("l", l_len);
    auto A = itensor::ITensor(i, j, k), B = itensor::ITensor(j, l, k);
    auto C = itensor::ITensor(i, l);
    auto D = alpha * A * B + beta * C;
\end{minted}
where alpha and beta are floating point variables.
CTF and TiledArray instead generate local index bindings (encoded by strings parsed at runtime) that encode elements of an expression. This is necessary, for example, to be able to use the same tensor with different index bindings in multiple parts of an expression or in multiple expressions. E.g., encoding of \cref{eq:example_DABC} in CTF,
\begin{minted}[frame=lines, linenos]{cpp} 
    Tensor<double> A(i_len, j_len, k_len), B(j_len, l_len, k_len);
    Tensor<double> C(i_len, l_len), D(i_len, l_len);
    D["il"] = alpha * A["ijk"] * B["jlk"] + beta * C["il"];
\end{minted}
and TiledArray, are remarkably similar:
\begin{minted}[frame=lines, linenos]{cpp} 
    TA::TiledRange shape_A{tr_i, tr_j, tr_k},  shape_B{tr_j, tr_l, tr_k};
    TA::TiledRange shape_C{tr_i, tr_l};
    TA::TArrayD A(world, shape_A),  B(world, shape_B);
    TA::TArrayD C(world, shape_C),  D(world, shape_C);
    D("i,l") = alpha * A("i,j,k") * B("j,l,k") + beta * C("i,l");
\end{minted}
where 1-d ranges \texttt{tr\_i}\ldots\texttt{tr\_l} , optionally tiled (e.g., to support block sparsity), are defined in terms of \texttt{i\_len}\ldots\texttt{l\_len} as:
\begin{minted}[frame=lines, linenos]{cpp} 
    TA::TiledRange1 tr_i{0, [...] i_len};
\end{minted}
 libtensor\cite{Ibrahim2014} provides a similar interface except that indices are independent objects, although not permanently bound as in ITensor. An example encoding of \cref{eq:example_DABC} in libtensor is,
\begin{minted}[frame=lines, linenos]{cpp}
    bispace<1> i_sp(i_len), j_sp(j_len), k_sp(k_len), l_sp(l_len);
    btensor<3> A(i_sp & j_sp & k_sp), B(j_sp & l_sp & k_sp);
    btensor<2> C(i_sp & l_sp), D(i_sp & l_sp);
    letter i, j, k, l;
    D(i|l) = alpha * mult(A(i|j|k), B(j|l|k)) + beta * C(i|l);
\end{minted}

NumPy \texttt{einsum} is probably the most widely adopted high-level interface. Conceptually similar to the interface of Cyclops CTF, it is based on concise, but highly expressive strings:
\begin{minted}[frame=lines, linenos]{python} 
    A = np.random.rand(i_len, j_len, k_len)
    B = np.random.rand(j_len, l_len, k_len)
    C = np.random.rand(i_len, l_len)
    D = alpha * np.einsum('ijk,jlk->il', A, B) + beta * C
\end{minted}
It should be noted that, in addition to their original eDSLs, many libraries such as CTF and TiledArray later included support for \texttt{einsum}, although these are typically layered on top of the lower-level eDSL. An important difference is that \texttt{einsum} supports an almost arbitrary operation string which could indicate contraction (of any of the cases discussed in \Cref{sec:preliminaries:operations}), as well as binary, unary, or tensor network (many-tensor) operations. In TAPP, the aim of a low-level interface which enables the highest level of optimization suggests targeting specific sub-classes of operations possible with \texttt{einsum} which can then be composed to accomplish more complicated operations.

\subsection{Interface Components}

\begin{figure}
    \centering
    \includegraphics[width=\textwidth]{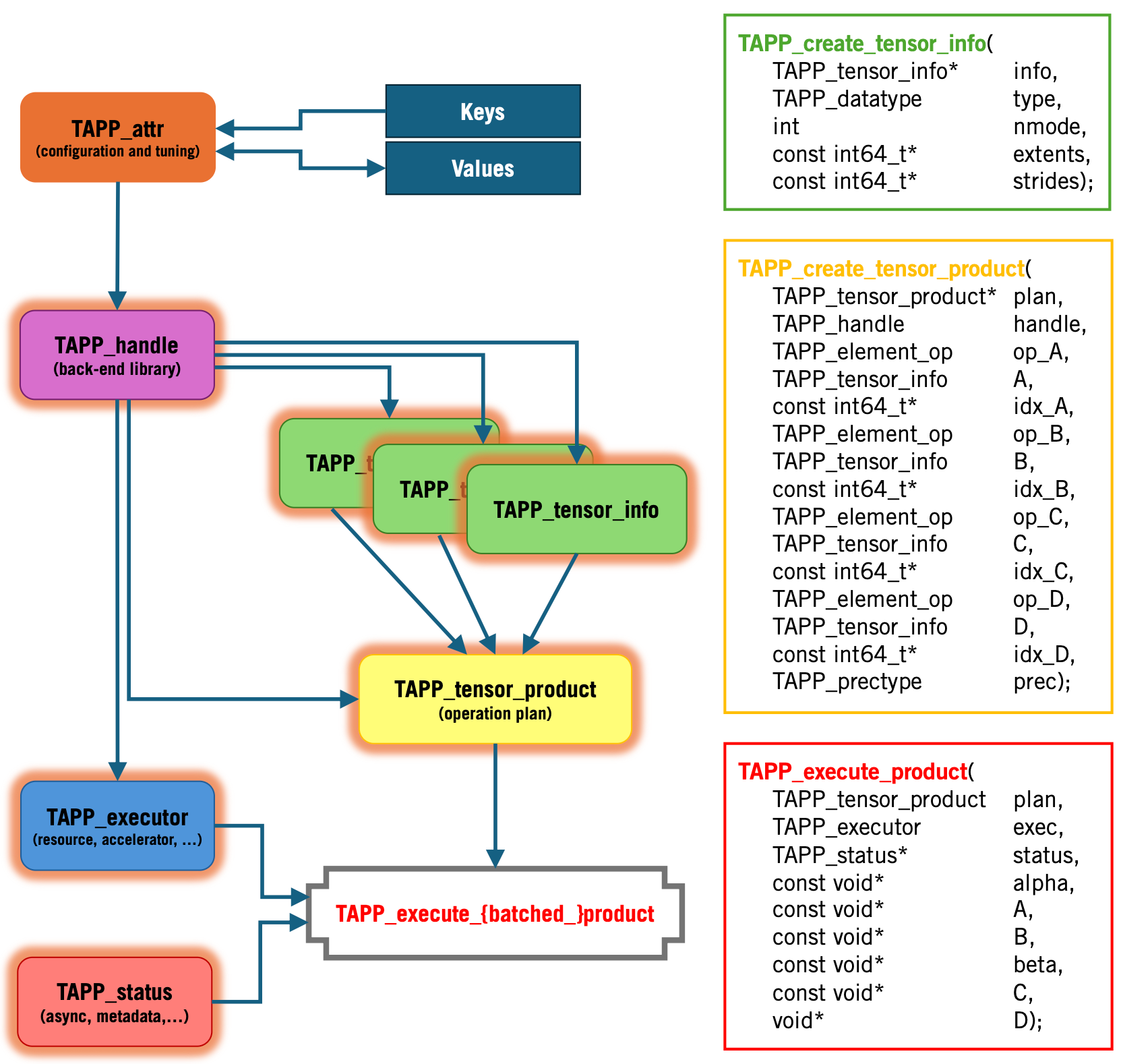}
    \caption{Flowchart of the TAPP interface components, with examples of TAPP API signatures. In addition to the dedicated \texttt{TAPP\_attr} object, all TAPP objects support key-value semantics, indicated by an orange glow.}
  \label{Fig:flowchart}
\end{figure}

The TAPP interface comprises a family of related data structures, represented as opaque object handles, an API for creating and manipulating objects, and finally a computational API which performs the desired computation. These elements are discussed in the rough order of the general workflow depicted in \Cref{Fig:flowchart}, which illustrates the relationship between the various objects and the computational API, along with example API signatures.

\subsubsection{The Library Handle} \label{sec:handle}

All TAPP operations, besides creating a bare key-value store, require a library handle. This object represents the internal state of the back-end, which may or may not be global. While not required, implementations are encouraged to save user resources by performing any expensive initialization when creating the library handle and then cleaning up when the handle is released. In more complex situations, multiple library handles obtained by the user may represent distinct TAPP back-ends with different (and incompatible) representations of other TAPP objects, algorithms, or other properties. Thus, each distinct library handle must be kept separate along with all objects created using that handle. On the other end of the spectrum, simpler back-ends may only define a single library handle as a constant, representing the global back-end state.

\subsubsection{Executors} \label{sec:executor}

An executor object can be obtained using the library handle, and represents one or more resources which are capable of performing tensor operations. For example, a CPU-based implementation may provide one or more executors corresponding to various partitions of the (physical or logical) CPU cores, to specific core bitmasks, to NUMA domains, etc. For a GPU-based back-end, executors could correspond to physical devices and/or to abstract concepts such as streams. For ease of use, libraries should typically define a constant ``default'' executor value which can be passed to other TAPP functions without initialization (similar to CUDA's ``stream 0'' or MPI's {\tt MPI\_COMM\_WORLD}).

\subsubsection{Tensor Descriptors} \label{sec:descriptor}

The tensor descriptor provides information about the logical shape and physical memory layout of a tensor. As a dense tensor library, TAPP specifies a \emph{general strided} layout, similar to {\tt std::layout\_stride} in C++23. The tensor descriptor also defines the data type of the tensor. However, it does \emph{not} define the pointer to tensor data. Rather, a tensor descriptor is an abstract specification of any tensor of the same shape and structure. This property allows the user additional flexibility through the ability to re-use tensor descriptors for similar data or to executed batched tensor operations.

\subsubsection{Tensor Operation Descriptors} \label{sec:operations}

As with the tensor descriptor, the abstract specification of a particular tensor operation is independent of the particular data and specific instance of execution of that operation. This abstract specification is stored in the tensor operation descriptor objects (one type for each distinct tensor operation), which are constructed based on a library handle and set of tensor descriptors, along with any other operation-specific information such as the desired computational precision. Separating the tensor operation descriptor from the actual operation execution, along with providing the tensor descriptors when initializing the operation rather than upon execution, is an important step to enable a range of optimization strategies in the library back-end. For example, the back-end can examine the operation and operands and heuristically (or even empirically) determine the optimal algorithm to use. Back-ends could even craft specific optimized algorithms using just-in-time compilation (JIT) technology. While providing an opportunity to allow such optimizations to occur, splitting the tensor operation descriptor from the execution also allows for the cost of performing such optimizations to be amortized over many executions.

\subsubsection{Execution}

Once the library handle, executor, and tensor operation descriptor (via the individual tensor descriptors) are obtained, an operation can be executed on specific data. At this point, only the pointers to the tensor data are required (along with scalars). Since the data type of the tensors are separately specified in the tensor descriptors, the execution API is naturally type-agnostic (although mixed-type support depends on the capabilities of the back-end). The same tensor operation description and other objects may be re-used for subsequent executions using the same or different data, encouraging users to cache or otherwise store these objects rather then creating them anew for each individual execution of an operation.

\subsubsection{Status Objects}

The execution of a tensor operation optionally results in the creating of a status object. This object can serve multiple purposes, and no single meaning is required by the TAPP interface. For example, in an asynchronous setting, the status object could provide a handle for waiting on completion, specifying data dependencies, etc. The status object could also provide metadata of the execution such as performance counters, execution time, or algorithmic details. The status object is distinct from the error code though (see \cref{sec:error-handling}), although it could potentially provide more data related to a reported error.

\subsubsection{Virtual Key-Value Stores (VKVs)}

The TAPP interface is intended to provide adaptability to a wide range of tensor back-ends, with varying options for tweaking performance and other settings. Likewise, the TAPP interface itself is expected to evolve over time as new architectures, implementation techniques, and algorithms emerge. Thus, TAPP includes a free-form mechanism for providing information to the underlying implementation through virtual key-value stores. Essentially, the key-value store represents a mapping from integral keys (e.g. an enumeration) to arbitrary binary values. These stores are called ``virtual'' because while they conform to the set/get API of a traditional map data structure, implementations are free to implement them in other ways, for example as a layer to access configuration values distributed throughout the library.

While TAPP does not currently define any required keys that back-ends must support, a number of use cases could be supported through this mechanism:
\begin{itemize}
\item Customized initialization of the library handle (see \cref{sec:handle}), or selection of a particular back-end through a wrapper interface.
\item Selecting executors (see \cref{sec:executor}) based on CPU core bitmasks, device IDs, or other criteria.
\item Providing information about data locality in tensor descriptors (see \cref{sec:descriptor}), for example NUMA domain, device memory, shared/pinned memory, etc.
\item Selection of particular algorithms or providing optimization hints in tensor operations (see \cref{sec:operations}).
\end{itemize}
TAPP provides a ``bare'' VKV object type, but unless otherwise noted all other TAPP objects also support setting and getting key-value pairs via a unified interface.

\subsubsection{Error Handling} \label{sec:error-handling}

Almost all TAPP API functions return an integral error code. Specific integer values for particular errors are not defined, but a plain text description of the error code is available, in analogy to {\tt strerror} in POSIX. Likewise, the user is always able to at least check for success. Because each function returns an error code, any output values are provided as pointers. In some cases, output values may be ignored (or rather, requested to not be returned) by passing {\tt NULL}. In some other cases, what would logically be purely input operands are instead treated as input/output and passed via pointer for security or safety reasons. For example, when TAPP objects are destroyed, the object handle is passed by pointer so that it can be set to a well-defined value indicating an uninitialized object (e.g. {\tt NULL}).

\subsubsection{Extensibility and Compatibility} \label{sec:extensibility}

The TAPP API is designed to be both forward and backward compatible. A major aspect of this is the virtual key-value store mechanism which allows not only implementation-specific behaviour, but also the ability to add standardized sets of keys as TAPP grows and evolves. As with other major standards such as MPI, C++, and OpenGL, extensions to TAPP will likely follow and codify common implementation-specific features that emerge either through convergence evolution or community consensus. In addition to standardizing key-value metadata, TAPP is intended to grow through the inclusion of new tensor operations, physical representations, or more complex internal structures through the addition of new API functions and/or TAPP objects. TAPP will rely heavily on community input and established practice to identify and define new functionality.


\section{The Reference Implementation}

The reference implementation is meant to provide a reliable and easy to read implementation for all the operations currently supported by TAPP. The main concern is therefore correctness and simplicity, while performance aspects are disregarded. 
For implementation developers, the reference implementation should be used as an example of what to implement when supporting TAPP. 
Other implementations can be more restrictive but should still yield the same result for the same input, unless that kind of input is not supported by the implementation. For example, implementations could disallow negative strides for technical or performance reasons, while in general implementations are expected to support far fewer mixed datatype combinations than the reference implementation, which supports arbitrary mixing of real and complex floating point numbers with 64, 32, and 16 (if compiler supported) bit widths. The reference implementation is developed in C like the interface, keeping it low-level, simple, and portable.


\subsection{Algorithm Structure}

\begin{figure}
\begin{minted}[frame=lines, linenos, escapeinside=||]{python} 
    function contract(alpha, beta, P, FA, FB, H, A, B, C) {
        size_FA <- prod(extent(i) for i in FA)
        size_FB <- prod(extent(j) for j in FB)
        size_P <- prod(extent(k) for k in P)
        size_H <- prod(extent(l) for l in H)

        L <- (0 for l in H)
        for (l in 0 to size_H-1) {
            I <- (0 for i in FA)
            for (i in 0 to size_FA-1) {
                J <- (0 for j in FB)
                for (j in 0 to size_FB-1) {
                    off_D <- offset(L, stride(D, l) for l in H) +
                            offset(I, stride(D, i) for i in FA) +
                            offset(J, stride(D, j) for j in FB)
                    D[off_D] <- beta * C[off_D]
                    
                    K <- (0 for k in P)
                    for (k in 0 to size_P-1) {
                        off_A <- offset(L, stride(A, l) for l in H) +
                                offset(I, stride(A, i) for i in FA) +
                                offset(K, stride(A, k) for k in P)
                        off_B <- offset(L, stride(B, l) for l in H) +
                                offset(J, stride(B, j) for j in FB) +
                                offset(K, stride(B, k) for k in P)
                        D[off_D] <- D[off_D] + alpha * A[off_A] * B[off_B]
                
                        increment(K, extent(k) for k in P)
                    } 
                    increment(J, extent(j) for j in FB)
                }   
                increment(I, extent(i) for i in FA)
            }
            increment(L, extent(l) for l in H)
        }
    }
\end{minted}
\caption{Pseudo-code for the implementation of general tensor contraction (Case 3). Decomposition of the index strings into subsets $\mathcal{P}$, $\mathcal{F_A}$, \ldots\ and implementation of basic functions such as \texttt{extent} and \texttt{stride} not shown.}
\label{Fig:implementation}
\end{figure}

The structure of the contraction algorithm starts by exposing the elements from the plan and tensor info objects such as indices, strides, and extents. With the core arguments exposed, the algorithm continues with error checking, which validates all of the prerequisites such as identical extents for indices sharing a label.

Before the main part of the algorithm (\Cref{Fig:implementation}), there is some necessary pre-processing to be done. In this pre-processing stage, indices are divided into Hadamard, contracted, and various free groups as outlined in \Cref{sec:preliminaries:operations}. Using the extracted indices, the loop sizes (product of involved extents) are calculated, and the strides and extents are extracted for each tensor and relevant index groups. While \Cref{Fig:implementation} shows distinct loops for each group of free/Hadamard indices, the implementation groups all indices in $D$ together (as $\mathcal{D}=\mathcal{H}\cup\mathcal{F_A}\cup\mathcal{F_B}$) in one loop structure. Repeated indices are processed by removing the repetitions and combining their strides. Stride arrays are then allocated for each tensor, with all tensors having a free index stride array and $A$ and $B$ additionally receiving contracted index strides. Strides that correspond to indices that don't appear in a particular tensor (e.g. $\mathcal{F_A}$ indices in $B$) are set to zero. 

The core of the algorithm then consists logically of four main loops, pictured in \Cref{Fig:implementation}. The first (outer) three loops iterate through combinations of indices which affect the location of values in all tensors $A$, $B$, and $D$ (and hence $C$ as they are assumed to have identical structure). The inner loop then iterates through combinations of indices affecting only $A$ and $B$, which produces values that can be directly added to the value of a single element of $D$, which is first initialized to a scaled value from $C$. Incrementing to the next combination of indices (here in reverse lexicographical order) and determining the actual position of tensor values in memory are handled by the \texttt{increment} and \texttt{offset} helper functions, respectively, as outlined in \Cref{Fig:helper-functions}. The actual implementation code only differs slightly, as mentioned above, and also by applying several obvious optimizations. Additionally, the reference implementation supports reduction over the input tensors $A$ and $B$ (not pictured in \Cref{Fig:implementation}) which constitutes an additional loop each for $A$ and $B$ to accumulate reduced values before multiplication. Finally, multiple data types and their mixtures are handled through a type tagging and basic helper functions (multiply, add, zero, etc.) which exhaustively implement the necessary combinations.

\begin{figure}
\begin{minted}[frame=lines, linenos, escapeinside=||]{python} 
    function increment(indices, shape) {
        for (k in 0 to len(shape)-1) {
            indices[k] <- (indices[k] + 1) % shape[k]
            if (indices[k] > 0) return
        }
    }
    
    function offset(indices, strides) {
        offset = 0
        for (k in 0 to len(strides)-1) {
            offset <- offset + indices[k] * strides[k]
        }
    }
\end{minted}
\caption{Pseudo-code for helper functions for the implementation of general tensor contraction.}
\label{Fig:helper-functions}
\end{figure}

\subsection{Operation Coverage}
The reference implementation covers the operations described in \cref{sec:preliminaries:operations} through Case~4, while only Case~3 is required by external TAPP implementations and depicted in \Cref{Fig:implementation}. The addition of Case~4 to the reference implementation was included for future-proofing, in case operation coverage is expanded in the future, and due to the relative ease with which it can be handled. As the implementation evolves, Case~5 will likely be added as well for completeness. Case~1, described in \cref{sec:preliminaries:operations:case1}, is covered by the inner three loops in the implementation, corresponding precisely to the well-known triple loop implementation of matrix multiplication. In this case, an index that is a part of tensor $D$ is a part of exclusively one of tensors $A$ and $B$. Iterating over the space of tensor $D$ affects the location of elements in exactly one of tensor $A$ and $B$ at a time. Case~2, described in \cref{sec:preliminaries:operations:case2}, introduces indices that are shared between all three tensors. These indices are processed as Hadamard products in the outer loop of the algorithm, and essentially creates a sequence of independent contractions of the type in Case~1. The algorithm allows this by letting iterations over the space of tensor D cause iterations in both tensor A and B simultaneously, which is in turn simply encoded in the stride arrays for each tensor (zero meaning that an index does not affect the location of elements in that tensor). As briefly mentioned above, Case~3 (\cref{sec:preliminaries:operations:case3}) is covered by preprocessing as well, where repeated indices are merged by summing their strides, effectively reducing them to a single ``simple'' index as in Cases~1 and 2. Cases~4 and 5 simply require additional loops, although as noted in \cref{sec:preliminaries:operations:case5}, doing so in a high-performance implementation can require either unacceptable additional workspace or floating point operations, which is why TAPP does not require implementations to support them.

\subsection{Challenge: Datatypes}
One challenge that came up during the development of the implementation was the handling of different datatypes. Allowing for input of different datatypes and mixed datatypes proved to be challenging in C, as there is no other way of doing it than specific code for each set of datatypes. More challenges that arose were the implementation of datatypes that C does not have native support for, such as 16 bit brain float and 16 bit float, which still are not implemented.

\subsection{Design Decisions}
During the development of the implementation, design decisions were made about the standard. Such design decisions included specifying what the operation will include, allowing for repeated and isolated indices. Furthermore, decisions around what kind of tensors to allow was made, allowing tensors with negative strides and zero mode tensors, scalars.

\subsection{Testing}
For the purpose of showing correctness of the reference implementation, a set of tests has been developed testing it against TBLIS. The tests uses semi-random tensors, where the number of modes, size, and values are randomized in a way that still allows for the purpose of each test.

Some tests worth mentioning is a test testing negative strides and a test testing strides with zero length. The features of these tests were particularly discussed, if the interface should have limits on what type of strides are allowed. Testing using tensors with zero modes is a special edge case because the tensor does not have any structural information as it essentially is a scalar.

For a full list of tests, see \cref{sec:testing}.

\section{Conclusion} 

We have introduced a detailed draft for the standard interface for tensor contraction operations, bridging differences between application fields and previously disconnected software. We defined a common terminology, provided the necessary software and established a decision-making committee to maintain the standard. The interface was integrated with several high-performance implementations and its use demonstrated within application software. 

\subsection{Software Integration}
As a proof-of-concept, bindings between TAPP and several tensor libraries have been successfully developed. TBLIS is the first library to officially support the TAPP interface directly within its own repository. Additionally, bindings for cuTENSOR have been provided in the TAPP repository, developed independently of the vendor. Finally, experimental bindings based on a generalization of TAPP\cite{Brandejs2025} for distributed memory tensors have been created for Cyclops CTF.\cite{Solomonik2014}

The first major software suite to adopt TAPP is DIRAC,\cite{dirac} specifically, the massively parallel coupled cluster module\cite{Pototschnig2021} has been updated to utilize TAPP as a standardized abstraction layer,\cite{Brandejs2025} decoupling the application from specific backend dependencies and including the experimental support for distributed memory libraries. 

\subsection{Outlook and Future work}
TAPP does not seek to provide every feature that a high-level tensor code might want, instead it focuses on functionality necessary to effectively encapsulate the performance-critical operations provided by tensor libraries.
To provide an example, one can reshape a multidimensional array formally without changing the way its elements are stored in memory. This we consider a convenience functionality, as it is not performance critical. In contrast to this, if one needs to perform such a transposition that shifts the array elements numbers in memory, one may need to call a high-performance library to do such reshuffle efficiently. TAPP aims to include the former and exclude the latter.

Despite the clear performance driven focus, there are two developments that are planned to follow the current work. We aim to provide a comprehensive set of randomized benchmarks, which will include representative problems from main application domains representing our tensor software community. For this we are collecting data on tensor sizes, structures and types of contractions. 

Related to the benchmark set, the second development is called Multi-TAPP, a feature allowing developers using TAPP to switch between tensor libraries on the runtime. This is similar to the concept of ``BLAS/LAPACK trampoline''\cite{kohler2013flexiblas,lapacktrampoline} published recently and allows seamless benchmarking and choosing dynamically the right tensor library to perform a given task.

\section*{Acknowledgements} 

J. Brandejs has received funding from the European Research Council (ERC) under the European Union's Horizon 2020 research and innovation programme (grant agreement No 101019907). Niklas H\"ornblad development of the reference implementation was supported by eSSENCE, the Swedish strategic research program in e-Science. EFV acknowleges support from the US National Science Foundation (award 2217081) and the US Department of Energy (award DE-SC0022327).
DAM acknowledges support from the US National Science Foundation (grant No CHE-2143725) and the US Department of Energy (grant No DE-SC0022893).

\begin{appendices}
\crefalias{section}{appendix}
\section{Glossary} \label{sec:glossary}
\begin{longtable}{p{1.25in} p{4.25in}}

\emph{Batch index} &
See \emph{Hadamard index}. \vspace{1em} \\

\emph{Broadcast} &
If a \emph{label} appears in an output tensor, but not in any input tensor, then elements will be repeated in the output data as many times as the \emph{extent} of that label. As a one-to-many process this is referred to as a \emph{broadcast}. Also referred to as: \emph{replication}. \vspace{1em} \\

\emph{Column-major layout} &
A particular \emph{strided layout} of a tensor in memory, where the the strides $\bigcup\limits_{s=1}^{n}\{s_k\}$ are given by $s_k=\prod\limits_{l=0}^{k-1}e_l$ and $\bigcup\limits_{s=1}^{n}\{e_k\}$ is the tensor \emph{shape}. \vspace{1em} \\

\emph{Contracted index} & An index whose label is shared between only input tensors. Also referred to as: \emph{internal index}. \vspace{1em} \\

\emph{Dimension} &
See \emph{extent}. \vspace{1em} \\

\emph{Extent} & The number of distinct integral values which an \emph{index} can represent, denoted as $d_i$ for the $i$th index or $\operatorname{extent}(i)$ for an index with label $i$. Typically a compact range such as in 0-based, $i \in [0,\operatorname{extent}(i))$, or 1-based, $i \in [1,\operatorname{extent}(i)]$ indexing. Also referred to as: \emph{length}, \emph{dimension}, \emph{range}, \emph{size} (although size can also refer to the product of all index extents, or equivalently the number of tensor elements). \vspace{1em} \\

\emph{External index} &
See \emph{Free index}. \vspace{1em} \\

\emph{Free index} & An index whose label is shared between an input and an output tensor. Also referred to as: \emph{external index}, \emph{uncontracted index}. \vspace{1em} \\

\emph{Hadamard index} &
An \emph{index} whose \emph{label} appears in all three tensors $A$, $B$, and $C$ in a ternary operation is referred to as a \emph{Hadamard index}. An operation with only Hadamard indices is referred to as a \emph{pure Hadamard contraction} or \emph{pure Hadamard product} (which is equivalent to an element-wise multiplication), while an operation with both Hadamard and non-Hadamard indices is a \emph{mixed Hadamard contraction} or \emph{mixed Hadamard product}. Also referred to as: \emph{batch} or \emph{batched index}. \vspace{1em} \\

\emph{Index} &
One of the $n$ labels representing either a range of integral values or one specific value. All $n$ indices together specify one or more elements of an $n$-index tensor. Also referred to as: \emph{mode} ($n$-\emph{mode} tensor), \emph{dimension} ($n$-\emph{dimensional} tensor, \emph{dimensionality} $n$, etc.), \emph{rank} (\emph{rank}-$n$ tensor), \emph{subspace}. In defining $n$, also $n$-\emph{adic} (for lower $n$, \emph{dyadic}, \emph{triadic}, etc.), $n$-\emph{way}, $n$-\emph{fold}, \emph{order}--$n$, and $n$-\emph{ary} tensor. Not to be confused with: \emph{label}. \vspace{1em} \\

\emph{Internal index} &
See \emph{contracted index} \vspace{1em} \\

\emph{Label} &
A particular symbol given to one or more of the $n$ indices of a tensor, representing either a range of integral values or one specific value. All $n$ indices together specify one or more elements of an $n$-index tensor. Also referred to as: \emph{index} (that is, the term \emph{index} is often used to refer both to the element of the tensor \emph{shape} and to a particular label assigned to that element). \vspace{1em} \\

\emph{Layout} &
See \emph{shape}. \vspace{1em} \\

\emph{Length} &
See \emph{extent}, \emph{shape}. \vspace{1em} \\

\emph{Mode} &
See \emph{index}. \vspace{1em} \\

\emph{$n$-adic, $n$-ary} &
See \emph{index}. \vspace{1em} \\

\emph{$n$-dimensional} &
See \emph{index}. \vspace{1em} \\

\emph{Order} &
See \emph{index}. \vspace{1em} \\

\emph{Permutation} &
See \emph{transpose}. \vspace{1em} \\

\emph{Range} &
See \emph{extent}. \vspace{1em} \\

\emph{Rank} &
See \emph{index}. \vspace{1em} \\

\emph{Reduction} &
If a \emph{label} appears in one input tensor, but not in any other tensor, then multiple elements (as many as the \emph{extent} of that label) of the tensor will be \emph{reduced} (summed) before multiplication by elements from the other input tensor in a ternary operation, or accumulation onto the output tensor in a binary operation. Also referred to as: \emph{summation} (although reduction may refer to binary operations other than addition). \vspace{1em} \\

\emph{Replication} &
See \emph{broadcast}. \vspace{1em} \\

\emph{Reshape} &
See \emph{transpose}. \vspace{1em} \\

\emph{Row-major layout} &
A particular \emph{strided layout} of a tensor in memory, where the the strides $\bigcup\limits_{s=1}^{n}\{s_k\}$ are given by $s_k=\prod\limits_{l=k+1}^{n-1}e_l$ and $\bigcup\limits_{s=1}^{n}\{e_k\}$ is the tensor \emph{shape}. \vspace{1em} \\

\emph{Shape} &
The set of \emph{extents} of an $n$-\emph{index} tensor. In abstract usage the set is unordered, reflecting potential permutations of the indices. \emph{Labels} assigned to the indices may refer to integral values within a range defined by the corresponding extent, while the shape as a whole determines the mathematical structure of the entire tensor. Also referred to as: \emph{size}, \emph{layout}, \emph{lengths}. \vspace{1em} \\

\emph{Size} &
See \emph{extent}, \emph{shape}. \vspace{1em} \\

\emph{Strided layout} &
A layout of a tensor in memory, where the memory location of an element with indices $\bigcup\limits_{k=1}^{n}\{i_k\}$ is given by $base + \sum\limits_{k=0}^{n-1} i_k\cdot s_k$ and $\bigcup\limits_{s=1}^{n}\{s_k\}$ are the tensor \emph{strides}. \vspace{1em} \\

\emph{Summation} &
See \emph{replication}. \vspace{1em} \\

\emph{Transpose} &
Two tensors with identical sets of labels but differing layouts of memory are said to be \emph{transposes} of each other. For typical layouts such as generalized column- or row-major order, this is equivalent to a change in the order of the labels. Also referred to as: \emph{reshape}, \emph{permutation}. \vspace{1em} \\

\emph{Uncontracted index} &
See \emph{Free index}. \vspace{1em} \\

\emph{Way} &
See \emph{index}. \vspace{1em}

\end{longtable}
\section{Testing} \label{sec:testing}
The test set for the reference implementation of TAPP contains the following test cases. Beyond the order of tests, the index strings themselves and their dimensions are randomized as well:
\begin{enumerate}
    \item Hadamard product
    \item Basic binary contractions
    \item Commutativity, swapping tensors and indices between A and B
    \item A check that changing around the indices for tensor D gives different permutations
    \item Contraction between tensors where all extents are the same size
    \item Outer product, no contracted indices
    \item Contractions where all indices are contracted
    \item Contractions where at least one tensor has zero modes
    \item Contractions where at least one tensor has one mode
    \item Contractions with sub-tensors, with the same number of modes as the larger tensors
    \item Contractions with sub-tensors, with a lower number of modes as the larger tensors
    \item Contractions with tensors using negative strides
    \item Contractions with sub-tensors using negative strides, and with the same number of modes as the larger tensors
    \item Contractions with sub-tensors using negative strides, and with a lower number of modes as the larger tensors
    \item Contractions with tensors using a mix of positive and negative strides
    \item Contractions with tensors using a mix of positive and negative strides, and with the same number of modes as the larger tensors
    \item Contractions with tensors using a mix of positive and negative strides, and with a lower number of modes as the larger tensors
    \item Contractions using double precision
    \item Contractions using complex numbers
    \item Contractions using complex numbers with double precision
    \item Contractions with a tensor with a zero stride
    \item Contractions with tensors having isolated indices, reductions
    \item Contractions with tensors having repeated indices
    \item Hadamard product combined with free indices
    \item Hadamard product combined with contracted indices
    \item Testing error handling for when extents for the same index do not match
    \item Testing error handling for when tensor C does not match tensor D
    \item Testing error handling for aliasing within tensor D
\end{enumerate}
\end{appendices}

\bibliography{references} 

\end{document}